\documentclass{iaus}
\usepackage{graphicx}

\newcommand{\hi}{\mbox{\rm H\,{\sc i}}}
\newcommand{\hii}{\mbox{\rm H\,{\sc ii}}}
\newcommand{\htwo}{\mbox{\rm H$_2$}}

\newcommand{\halpha}{\mbox{\rm H$\alpha$}}

\title[Scaling Relations Between Gas and Star Formation in Nearby Galaxies] 
{Scaling Relations between Gas and Star Formation in Nearby Galaxies}

\author[Frank Bigiel, Adam Leroy, Fabian Walter]   
{Frank Bigiel$^1$, Adam Leroy$^2$ \and Fabian Walter$^3$}

\affiliation{$^1$Department of Astronomy, Radio Astronomy Laboratory,
University of California, Berkeley, CA 94720, USA \\ email: {\tt bigiel@astro.berkeley.edu} \\[\affilskip]
$^2$National Radio Astronomy Observatory, 520 Edgemont Road, Charlottesville,
VA 22903, USA\\[\affilskip]
$^3$Max-Planck-Institut f{\"u}r Astronomie, K{\"o}nigstuhl 17,
69117 Heidelberg, Germany}

\pubyear{2010}
\volume{270}  
\pagerange{1--8}
\jname{Computational Star Formation}
\editors{Alves, Elmegreen, Girart, Trimble, eds.}
\begin{document}

\maketitle

\begin{abstract}

High resolution, multi-wavelength maps of a sizeable set of nearby
galaxies have made it possible to study how the surface densities of
\hi, \htwo\ and star formation rate ($\Sigma_{\rm HI}$,$\Sigma_{\rm
  H2}$,$\Sigma_{\rm SFR}$) relate on scales of a few hundred
parsecs. At these scales, individual galaxy disks are comfortably
resolved, making it possible to assess gas-SFR relations with respect to
environment within galaxies. $\Sigma_{\rm H2}$, traced by CO
intensity, shows a strong correlation with $\Sigma_{\rm SFR}$ and the
ratio between these two quantities, the molecular gas depletion
time, appears to be constant at about 2\,Gyr in large spiral
galaxies. Within the star-forming disks of galaxies, $\Sigma_{\rm
  SFR}$ shows almost no correlation with $\Sigma_{\rm HI}$. In the
outer parts of galaxies, however, $\Sigma_{\rm SFR}$ does scale
with $\Sigma_{\rm HI}$, though with large scatter. Combining data from
these different environments yields a distribution with multiple
regimes in $\Sigma_{\rm gas}-\Sigma_{\rm SFR}$ space. If the
underlying assumptions to convert observables to physical quantities
are matched, even combined datasets based on different SFR tracers,
methodologies and spatial scales occupy a well define locus in
$\Sigma_{\rm gas}-\Sigma_{\rm SFR}$ space.
  
\end{abstract}

\firstsection 
\section{Introduction and the Global Star Formation Law}


Great progress has been made towards an understanding of star formation
(SF) on small scales in the Milky Way, but many open questions remain
about its connection to large scale processes: what sets where SF
occurs in galaxies and how efficiently gas is converted into stars?
How important are global, galaxy-scale environmental parameters as
opposed to small-scale properties of the interstellar medium (ISM)?
What is the role of feedback in regulating SF? To address such
questions, theoretical modeling and simulations need to be constrained
by comprehensive observations.

Both observations and theory have focused on the relationship between
the star formation rate (SFR) and the gas density, for which a
tight power-law relationship was observed in
a large number of galaxies by \cite{kennicutt98}. Such a relationship
was first suggested many decades ago by \cite{schmidt59}, who studied
the distributions of atomic gas and stars in the Galaxy. Over the following
decades, similar studies targeted individual Local Group galaxies, e.g.,
M33 (\cite{madore74,newton80}), the Large Magellanic Cloud
(\cite{tosa75}), and the Small Magellanic Cloud (\cite{sanduleak69}).
\cite{kennicutt89} carried out the first comprehensive extragalactic
study targeting a large sample of nearby galaxies and
\cite{kennicutt98} followed up this work, focusing on measurements
averaged across galaxy disks.  In a sample of 97 nearby normal and
starburst galaxies, he found a close correlation between the
galaxy-average total gas surface density ($\Sigma_{\rm gas} =
\Sigma_{\rm HI} + \Sigma_{\rm H2}$) and the galaxy-average SFR surface
density ($\Sigma_{\rm SFR}$). Following this work, it has become
standard to study the relationship between gas and star formation via
surface densities, which are observationally more easily accessible than
volume densities.

\begin{figure}[t]
\begin{center}
\includegraphics[width=1.5in]{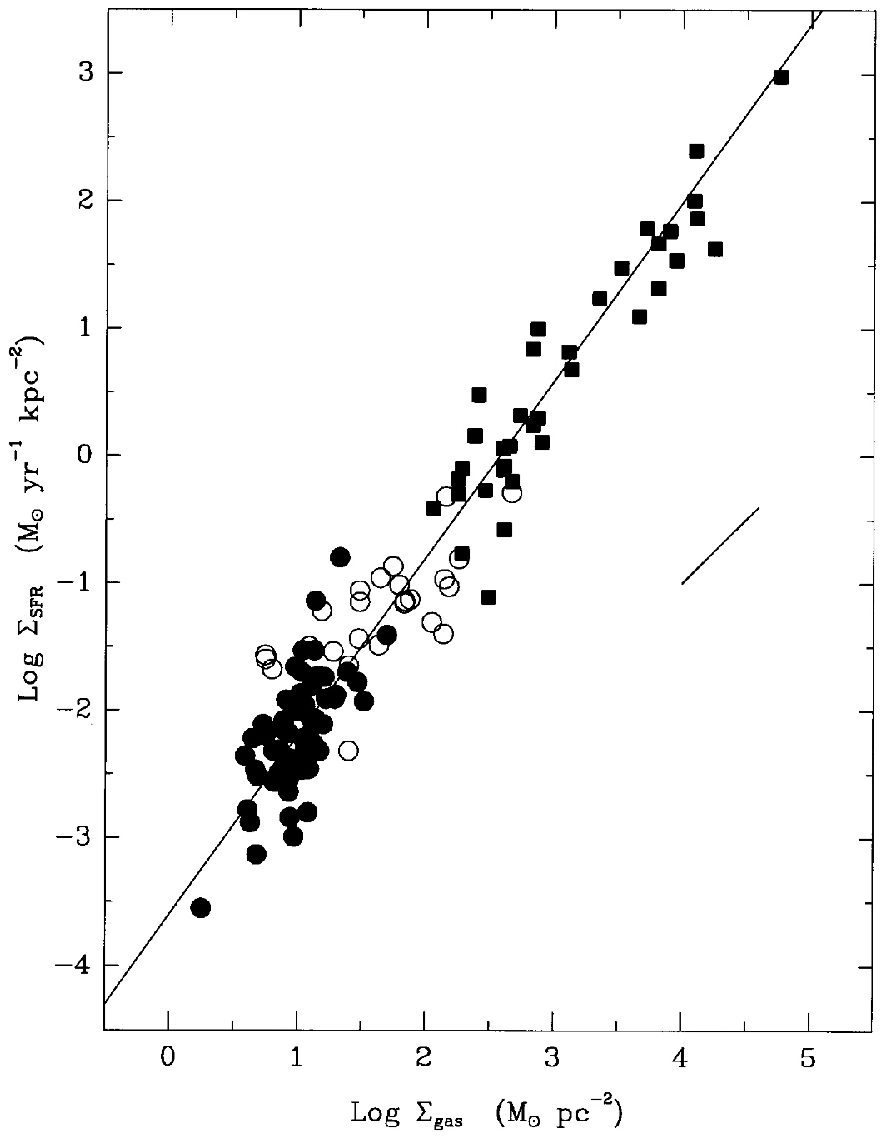} 
\includegraphics[width=1.5in]{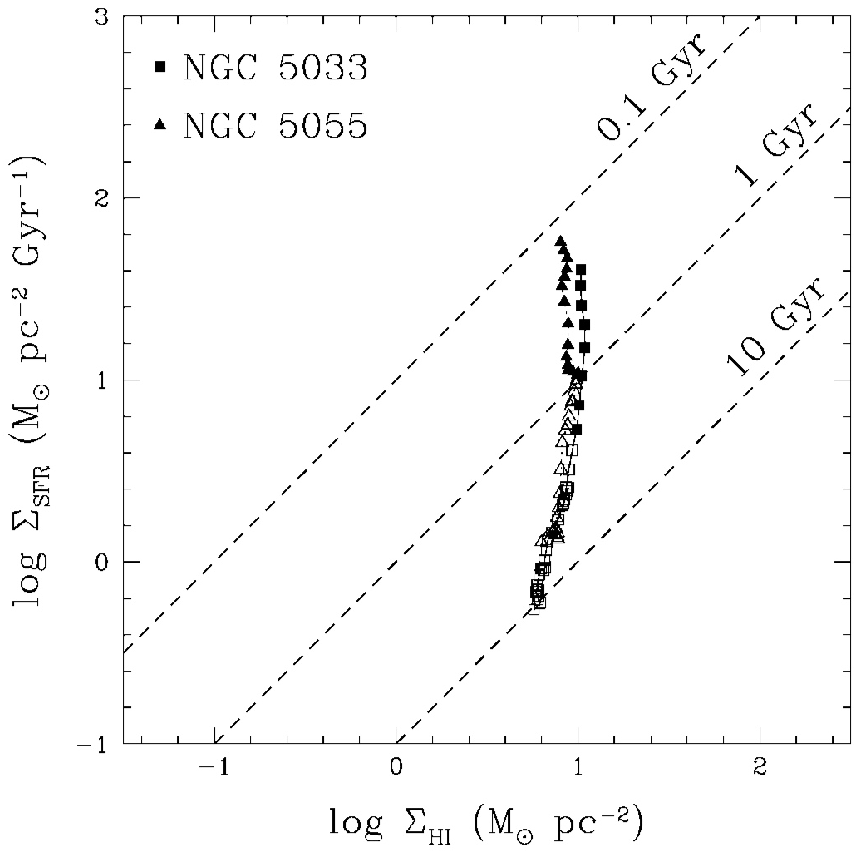} 
\includegraphics[width=1.5in]{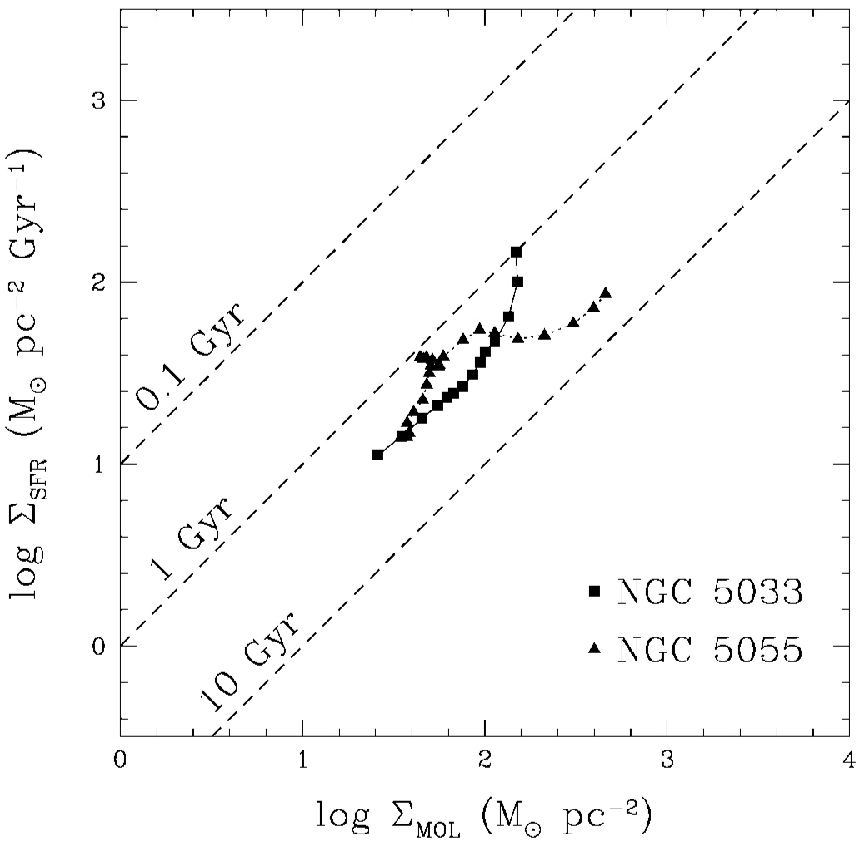} 
\caption{{\em Left:} \cite{kennicutt98} found a strong correlation
  over many orders-of-magnitude between global averages of
  $\Sigma_{\rm SFR}$ and $\Sigma_{\rm gas} = \Sigma_{\rm
    HI}+\Sigma_{\rm H2}$ for a large sample of nearby normal and
  starburst galaxies. He derived a power law index $N\approx1.40$,
  implying more efficient SF for galaxies with higher average gas
  columns. {\em Middle and Right:} Surface densities of gas and
  SFR measured in radial profiles by \cite{wong02} for two
  exemplary nearby spirals. The middle panel shows $\Sigma_{\rm SFR}$
  versus $\Sigma_{\rm HI}$, the right panel $\Sigma_{\rm SFR}$ versus
  $\Sigma_{\rm H2}$. \cite{wong02} found no correlation
  between atomic gas and SFR, whereas molecular gas and SFR scale with
  one another.}
\label{fig1}
\end{center}
\end{figure}

\cite{kennicutt98} found $\Sigma_{\rm SFR}=A\times\Sigma_{\rm
  gas}^{N}$, with intercept $A$ and power law index $N$ --- a
relationship that is variously referred to as the ``star formation
law,'' ``Schmidt-Kennicutt law,'' or ``Schmidt Law.''
\cite{kennicutt98} derived $N\approx1.40$. Because the ratio $\Sigma_{\rm SFR}/\Sigma_{\rm gas}$
describes how efficiently gas is converted into stars (and is thus often referred
to as the star formation efficiency, SFE), this super-linear power law index implies that systems with
higher average gas surface densities more efficiently convert gas into
stars (left panel, Figure \ref{fig1}). This measured value is close to $N=1.5$,
which is expected if the free-fall time in a fixed scale height gas
disk is the governing timescale for SF on large scales. Other studies
working with disk-averaged, global measurements found $N$ to be in the
range of $\sim0.9-1.7$ (e.g., \cite{buat89,buat92,deharveng94}).

The availability of high-resolution maps of CO emission, the standard
tracer of molecular gas, made it possible to follow up the work of
\cite{kennicutt98} with studies focusing on azimuthally-averaged
radial profiles of gas and SF. Resolving galaxies in this
way makes it possible to look at how gas and SF relate within
individual galaxy disks, opening up a wide range of environmental factors
to explore. \cite{wong02} used BIMA SONG data (\cite{helfer03}) to
study 6 nearby spirals, \cite{boissier03} explored a larger sample of
nearby spirals, \cite{heyer04} studied the Local Group galaxy M33, and
\cite{schuster07} explored the gas-SF relation in M51. These studies
derived power law indices in the range $N\approx1-3$, leaving it
unclear whether a single relation relates gas and SF when galaxy disks
are spatially resolved. Further disagreement centered on the relationship of SF to
different types of gas --- \hi , H$_2$, and total gas. Intuitively,
one might expect a stronger correlation between SF and the cold,
molecular phase, rather than the atomic phase. \cite{wong02} indeed
found a much stronger correlation of $\Sigma_{\rm SFR}$ with the
molecular gas, $\Sigma_{\rm H2}$ (compare Figure \ref{fig1}). However,
\cite{kennicutt98} and \cite{schuster07} both found a better
correlation of $\Sigma_{\rm SFR}$ with the total gas, $\Sigma_{\rm
  gas}$, than with $\Sigma_{\rm H2}$. 
  

\section{Recent Advances: Gas and Star Formation on sub-kiloparsec Scales}
\label{subkpc}

One reason that different studies returned such different results were the
wide range of SFR tracers employed in the various
analyses. Furthermore, different studies employed widely varying
methods to correct the observed UV and H$\alpha$ intensities for the
effects of extinction by dust. Because the correction factor is usually
$\gtrsim 2$, the adopted methodology makes a large
difference. A large step forward in this field came from the {\em
  Spitzer} space telescope. As part of SINGS ({\em Spitzer} Infrared
Nearby Galaxies Survey, \cite{kennicutt03}), {\em Spitzer} obtained
high-resolution IR maps of a large sample of nearby
galaxies. \cite{calzetti05} and \cite{calzetti07} demonstrated the utility
of these maps to trace recently formed stars obscured on small scales,
particularly when used in combination with H$\alpha$ --- a tracer of
unobscured star formation.

At the same time the VLA large program THINGS (\cite{walter08})
obtained the first large set of high resolution, high sensitivity
21-cm line maps for the same sample of galaxies. Following shortly
thereafter, the IRAM large program HERACLES mapped CO emission for an
overlapping sample of nearby galaxies (first maps in \cite{leroy09}). The result was, for the
first time, a matched set of sensitive, high spatial resolution maps
of atomic gas, molecular gas, embedded and unobscured star formation
for a large sample of nearby galaxies. The resolution of the maps
allowed hundreds of independent measurements per galaxy, leading to
significantly improved statistics and the ability to carefully isolate
regions with specific physical conditions.

\begin{figure}[t]
\begin{center}
\includegraphics[width=6in]{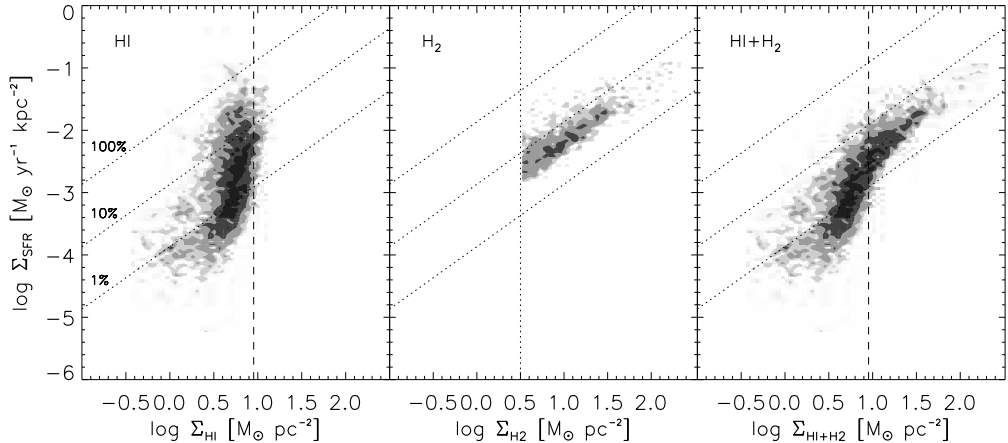} 
\caption{$\Sigma_{\rm SFR}$ versus $\Sigma_{\rm HI}$ (left),
  $\Sigma_{\rm H2}$ (middle) and $\Sigma_{\rm gas}$ (right panel) for
  pixel-by-pixel data from 7 nearby spirals at 750\,pc resolution. The
  contours represent the density of sampling points (pixels), where
  darker colors indicate a higher density. The \hi\ distribution
  saturates at about 10~M$_\odot$~pc$^{-2}$ and shows no correlation
  with the SFR (left panel). Gas in excess of this surface density is
  predominently molecular. The middle panel illustrates the
  correlation between \htwo\ and SFR, which can be described by a
  power law with slope $N\approx1.0$. This implies a constant
  \htwo\ depletion time of $\sim2$\,Gyr. The total gas plot (right
  panel) illustrates the different behavior of \hi\ and
  \htwo\ dominated ISM at low and high gas column densities,
  respectively.}
\label{fig2}
\end{center}
\end{figure}

\cite{bigiel08} combined these data to compare \hi, \htwo, and SFR
across a sample of 7 nearby spiral galaxies. Figure \ref{fig2} shows
the results of this analysis: the left panel shows $\Sigma_{\rm HI}$,
the middle panel $\Sigma_{\rm H2}$, and the right panel $\Sigma_{\rm
  gas}=\Sigma_{\rm HI}+\Sigma_{\rm H2}$ versus $\Sigma_{\rm SFR}$
(derived from a combination of far UV and 24$\mu$m emission). Galex far UV 
emission was chosen to trace the recent, unobscured SF because of the
low background in the FUV channel and the large field-of-view of the GALEX
satellite. In these
plots, \hi\ and \htwo\ show distinct behaviors: the atomic gas
shows no clear correlation with the SFR, whereas the molecular gas
exhibits a strong correlation. As a result, the composite total gas-SFR
relation is more complex than a single power law. In the combined
(total gas) plot in the right panel, one can clearly distinguish where
the ISM is \hi\ dominated (low gas columns, steep relation) form where
it is \htwo\ dominated (high gas columns, roughly linear correlation).

If a power law is fit to the molecular gas distribution in the middle
panel, one obtains $N\approx1.0$. This can be restated as a constant
ratio $\Sigma_{\rm SFR}/\Sigma_{\rm H2}$, which means that on average each parcel of \htwo\ forms
stars at the same rate. \cite{leroy08} searched for correlations
between $\Sigma_{\rm SFR}/\Sigma_{\rm H2}$ and a number of
environmental variables --- ISM pressure, dynamical time,
galactocentric radius, stellar and gas surface density --- and found
little or no variation across the disks of 12 nearby spirals. On the
other hand many of these same environemntal variables do correlate
strongly with the \htwo-to-\hi\ ratio. The combined conclusion of
these two studies was that the average depletion time in the molecular
gas (i.e., $\Sigma_{\rm H2}/\Sigma_{\rm SFR}$) of nearby spirals is fairly constant at $\sim2.0$\,Gyr, while the
abundance of molecular gas is a strong function of environment.

\cite{blanc09} carried out a similar experiment to \cite{bigiel08}.
They sampled the inner part of M51 with $170$\,pc diameter apertures
and estimated local SFR surface densities from
\halpha\ emission. Their integral field unit observations allowed for
accurate estimates of internal extinction and corrections for
contamination by the AGN and diffuse ionized gas. They used a
Monte-Carlo approach to incorporate upper limits into their power law
fit. Their results are in good agreement with \cite{bigiel08} regarding M51 in particular
as well as regarding the general conclusions they reached: a
virtual absence of correlation between SFR and \hi, a strong
correlation between SFR and \htwo\ and a molecular gas depletion time
that shows little variation with molecular gas column. 

Recently, \cite{rahman10} explored the impact of different SFR
tracers and the role of possible contributions from diffuse emission
and different sampling and fitting strategies on the relationship
between $\Sigma_{\rm H2}$ and $\Sigma_{\rm SFR}$. They found that the
SFR derived for low surface brightness regions is extremely sensitive
to the underlying assumptions, but that at high surface brightness the result of a
roughly constant \htwo\ depletion time is robust.

\begin{figure}[t]
\begin{center}
\includegraphics[width=2.5in]{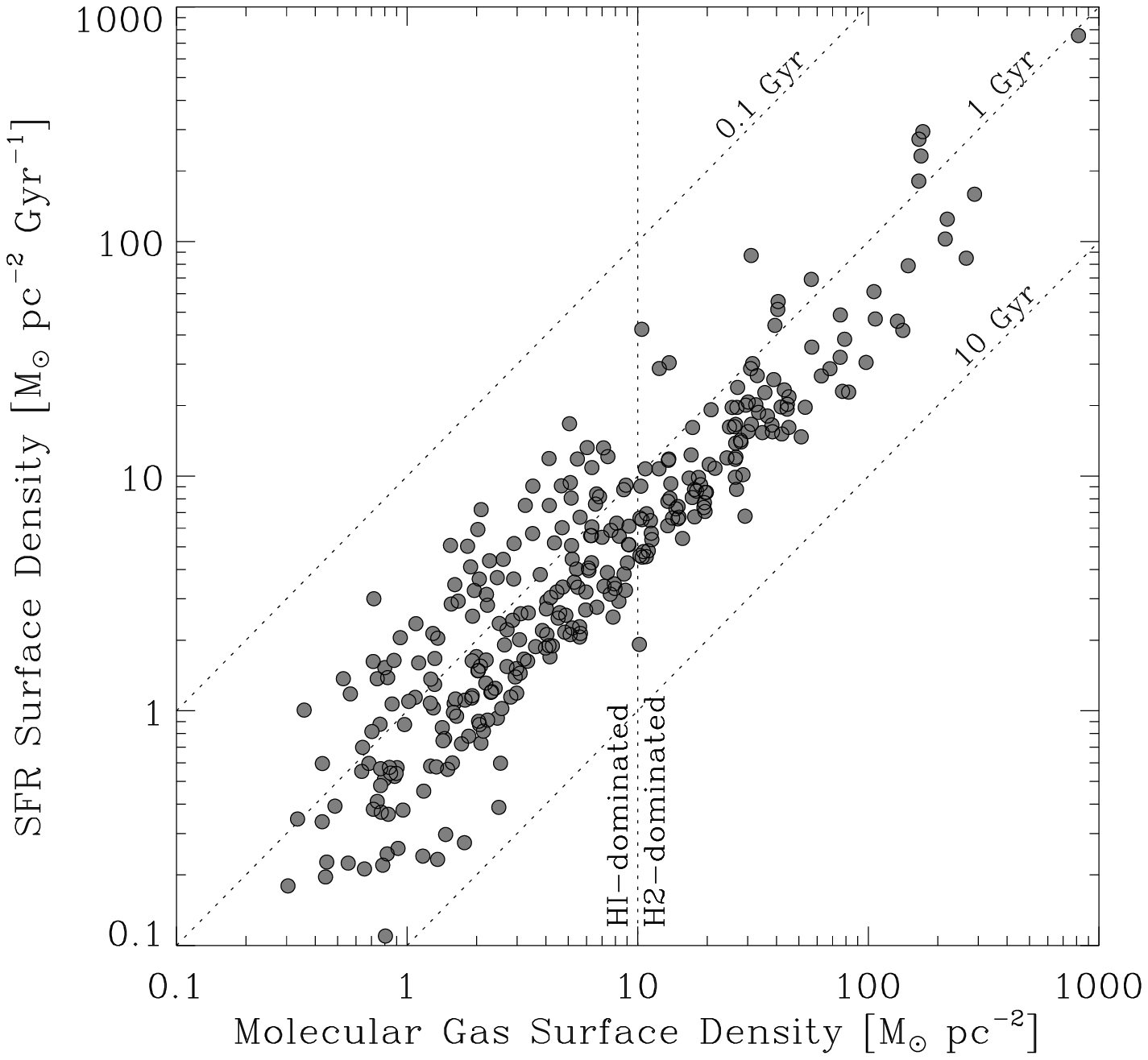} 
\includegraphics[width=2.5in]{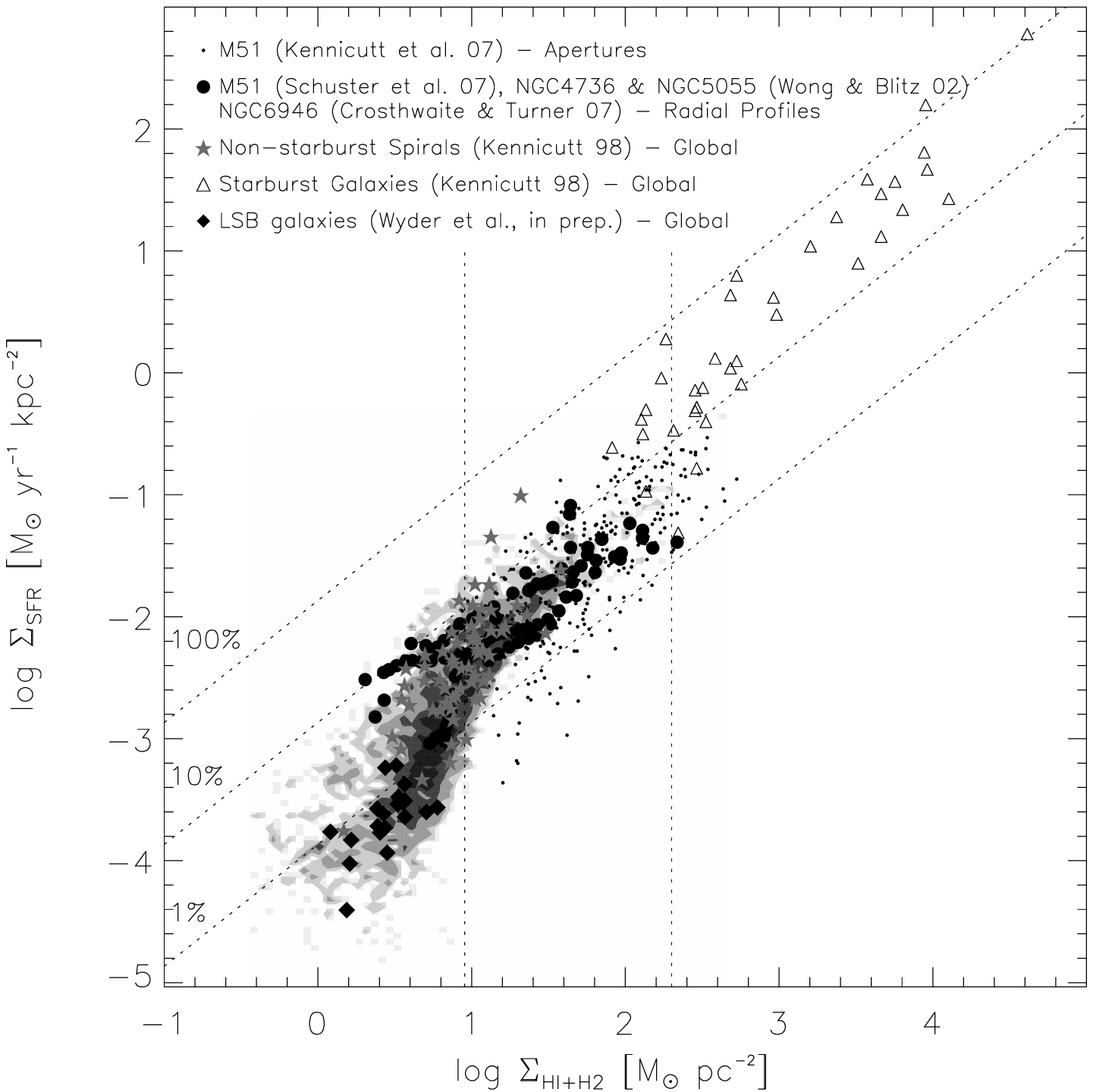} 
\caption{{\em Left:} $\Sigma_{\rm SFR}$ versus $\Sigma_{\rm H2}$ from
  Schruba et al. (in prep.). They apply a stacking analysis to the
  HERACLES CO data to probe the \htwo-SFR relation far into the regime
  where $\Sigma_{\rm HI} > \Sigma_{\rm H2}$. The correlation between
  $\Sigma_{\rm SFR}$ and $\Sigma_{\rm H2}$ extends smoothly into the
  \hi -dominated parts of galaxies out to 1.2\,r$_{25}$. {\em Right:} Comparison between
  different datasets using different methodologies from
  \cite{bigiel08}. The contours are identical to the right panel in
  Figure \ref{fig2} and the overplotted symbols come from studies
  using a variety of SFR tracers and methodologies. All datasets have
  been adjusted to match the same set of assumptions when
  converting observables to physical quantities. The composite sample occupies a
  well-defined locus in $\Sigma_{\rm SFR}-\Sigma_{\rm gas}$ space.}
\label{fig3}
\end{center}
\end{figure}

Even more recently, Schruba et al. (in prep.) combined the HERACLES
and THINGS data to coherently average CO spectra as a function of
radius. With this approach they are able to trace molecular gas out to
1.2\,r$_{25}$, allowing them to study the relation between \htwo\ and
SFR where \hi\ dominates the ISM. They demonstrate that the tight
correlation between \htwo\ and SFR crosses seamlessly into the \hi
-dominated outer disk (left panel Figure \ref{fig3}).

With so much effort expended measuring SFRs and gas densities over the years, it
is interesting to ask whether the literature largely agrees. The right
panel in Figure \ref{fig3} shows a collection of literature
measurements along with the data from \cite{bigiel08}. After
matching underlying assumptions about how to derive physical
quantities from the observables, the literature data populate a well-defined locus in
$\Sigma_{\rm SFR}$--$\Sigma_{\rm gas}$ space.

With some consensus emerging on the broad distribution of data in
SFR-\htwo\ space, attention is turning to the origin of the intrinsic
scatter in the SFR-\htwo\ ratio. \cite{schruba10} looked at this
as a function of spatial scale in M33 and showed that scatter in
the CO-to-H$\alpha$ ratio increases dramatically once a resolution element
contains only a single star-forming region (i.e., \hii\ region or giant
molecular cloud). This occurs at scales of $\sim 150$~pc in M33 but
should be a function of the environment studied. They interpreted this
finding as a result of the evolution of these regions, so that
information on the life cycle of giant molecular clouds is embedded in
the $\Sigma_{\rm SFR}$-$\Sigma_{\rm H2}$ relation, especially at high resolutions.

Another avenue of investigation is the role of the host galaxy in driving
the scatter in the SFR-\htwo\ ratio. \cite{leroy08} and \cite{bigiel08}
found little environmental dependence of this quantity in large spiral
galaxies, but with the completion of the HERACLES survey ($47$ galaxies
spanning from low-mass dwarfs to large spirals) we can now apply a
similar analysis to a much wider sample of galaxies. Bigiel et al. (in
prep.) and Schruba et al. (in prep.) explore how scatter in the
SFR-to-\htwo\ ratio breaks into scatter {\em within} galaxies and
scatter {\em among} galaxies. They find that scatter among galaxies
dominates the relation, with a clear trend evident so that less
massive, more metal poor, later type galaxies show systematically
higher ratios of SFR-to-\htwo. \cite{young96} saw this trend using
integrated FIR-to-CO ratios. These new analyses show that once these
galaxy-to-galaxy variations are removed, galaxies exhibit a very tight
internal \htwo-SFR relation.

\section{Scaling Relations beyond the Optical Disks}

\hi\ maps reveal atomic gas out to many optical radii in spiral
galaxies and over the past few years, GALEX UV observations have
revealed widespread SF in the outer parts of many galaxies (e.g.,
\cite{thilker05, gildepaz07,bigiel10b}). These outer disks have fewer
heavy elements, less dust, and lower stellar and gas surface densities
than the inner parts of spiral galaxies. Contrasting the gas-SFR
relationship in outer disks with that in the inner parts of normal galaxies gives a
chance to assess the impact of these parameters on SF.

\begin{figure}[t]
\begin{center}
\includegraphics[width=2.5in]{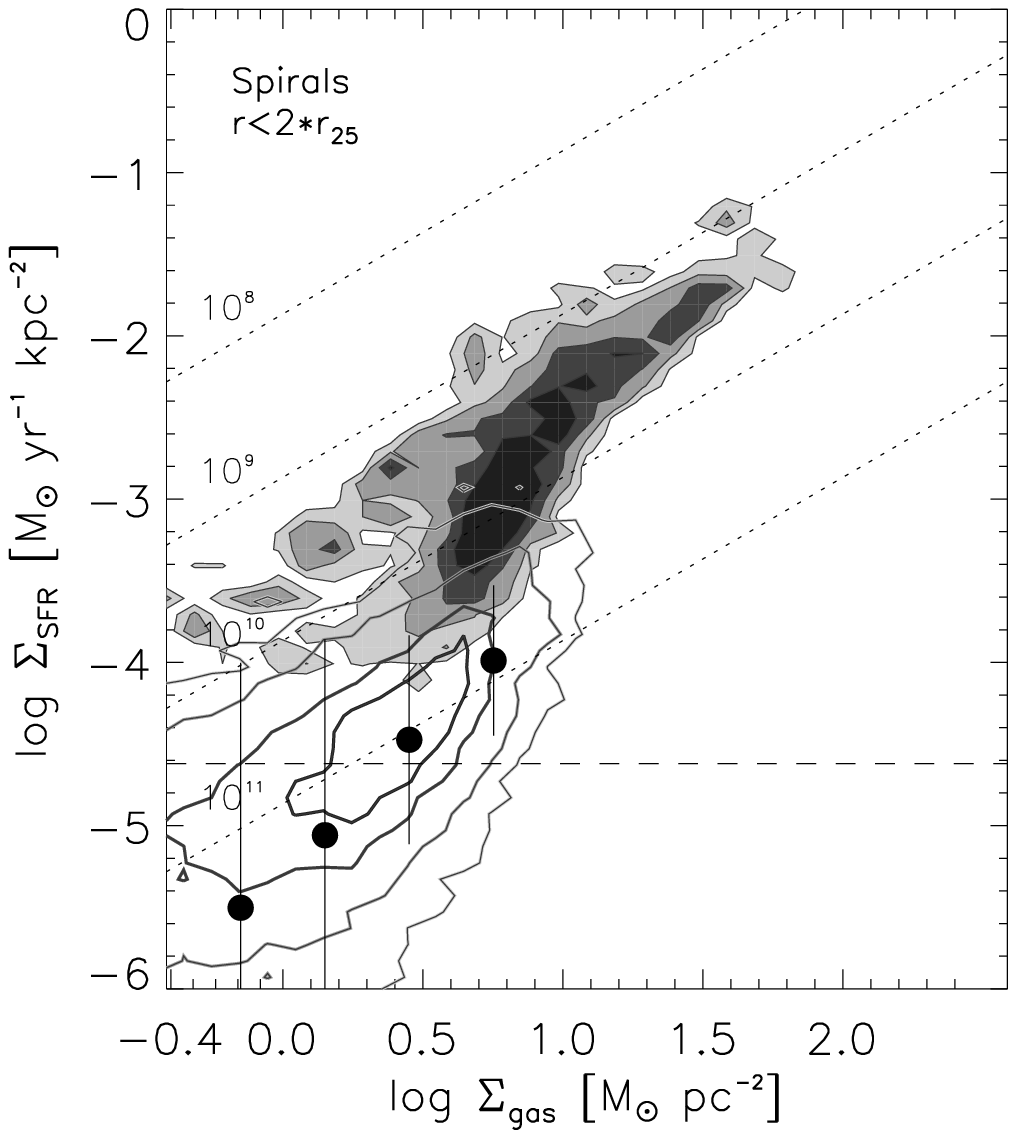} 
\includegraphics[width=2.5in]{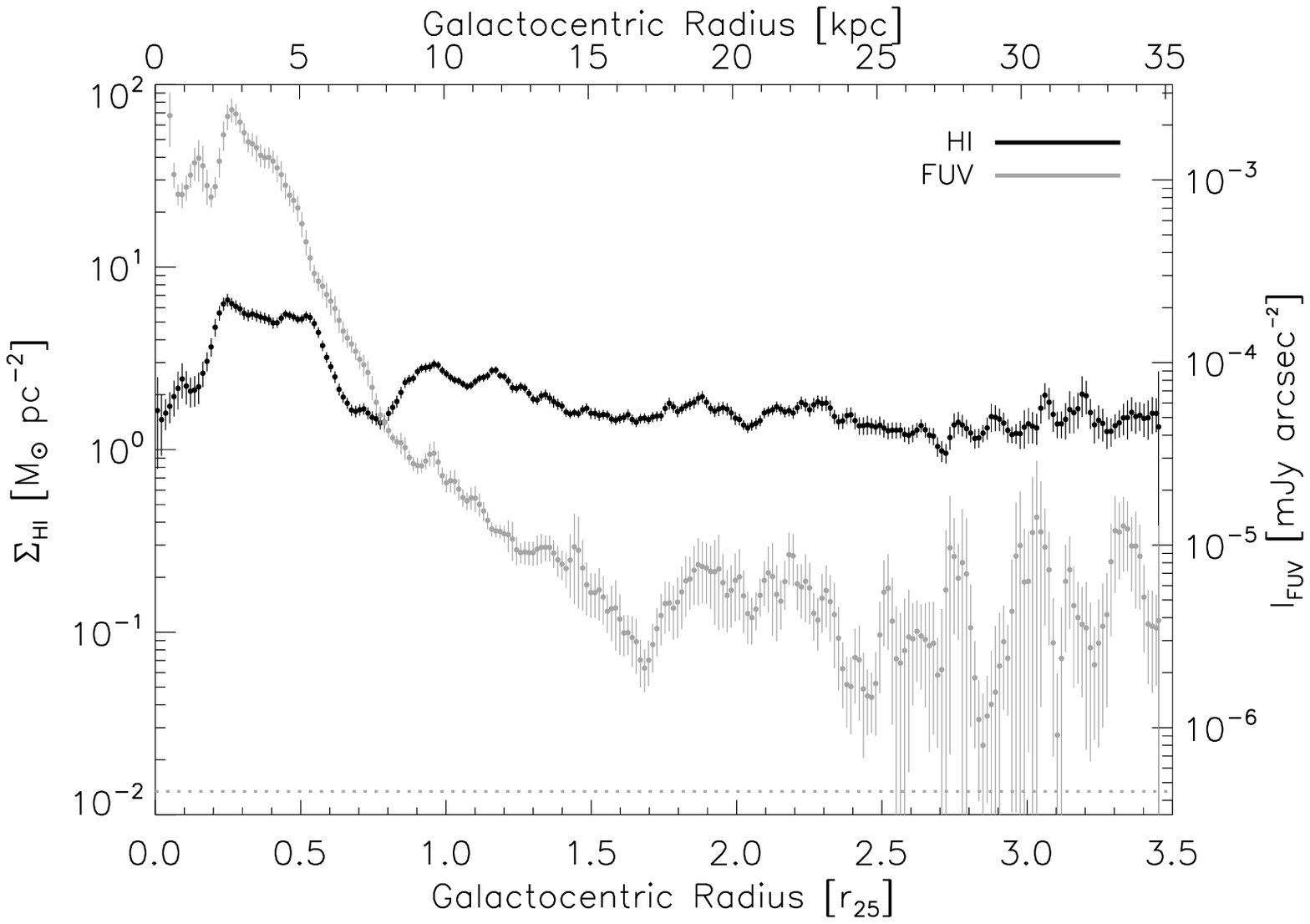} 
\caption{{\em Left:} $\Sigma_{\rm SFR}$ versus $\Sigma_{\rm gas}$ for
  the outer (open contours) and inner (filled contours) parts of
  nearby spiral galaxies (\cite{bigiel10a}). The combined distribution
  reveals multiple regimes: at large radii and low surface densities,
  $\Sigma_{\rm SFR}$ scales with $\Sigma_{\rm gas}\approx\Sigma_{\rm
    HI}$. Over most of the area in disks, $\Sigma_{\rm SFR}$ is a very
  steep function of $\Sigma_{\rm gas}$, with the \htwo-to-\hi\ ratio
  being the key determinant of $\Sigma_{\rm SFR}$. At high column densities,
  the gas is predominantly molecular and correlates well with
  $\Sigma_{\rm SFR}$. At even higher column densities, a steepening of
  this relation, meaning an increasing efficiency of SF, may accompany the
  transition from galaxy disks to starbursts. {\em Right:} \hi\ and
  far UV radial profiles for M83 out to almost 4 optical radii
  r$_{25}$. The inferred \hi\ depletion time is about a Hubble time at
  large radii, much longer than the molecular gas depletion times
  measured in many nearby spirals.}
\label{fig4}
\end{center}
\end{figure}

In the left panel of Figure \ref{fig4} we show the results of a
pixel-by-pixel analysis of outer galaxy disks: the open contours show
$\Sigma_{\rm SFR}$ versus $\Sigma_{\rm gas}$ for a sample of 17 spiral
galaxies at 15" resolution (corresponding to physical scales between
200\,pc and 1\,kpc) from \cite{bigiel10a}. SFRs are estimated from
GALEX far UV emission and $\Sigma_{\rm gas}$ is estimated from HI
emission alone, assuming a negligible contribution from molecular gas
on $\sim$kpc scales in outer galaxy disks. For comparison, the filled
contours show sampling data from the star forming disks of 7 spiral
galaxies from \cite{bigiel08} (compare the right panel in Figure
\ref{fig2} above). In the outer disks (open contours) $\Sigma_{\rm
  SFR}$ scales with $\Sigma_{\rm gas}$, i.e., $\Sigma_{\rm HI}$,
though the scatter in $\Sigma_{\rm SFR}$ for a given \hi\ column is
large. This is an interesting difference compared to the inner
parts of spiral galaxies, where the \hi\ showed no clear correlation
with $\Sigma_{\rm SFR}$ (compare Section \ref{subkpc}). 

The \hi -FUV relation observed for outer disks suggests two
things. First, that at large radii the availability of \hi\ may be a
bottleneck for star formation. Even if stars form directly from \htwo,
molecular clouds must be assembled from \hi\ and this will only be possible
in regions with enough \hi\ to assemble these clouds. Second, in the inner parts of
galaxies many physical conditions important to the \hi-\htwo\
conversion change while $\Sigma_{\rm HI}$ remains approximately fixed,
but in the outer parts $\Sigma_{\rm HI}$ varies while other
environmental conditions show comparatively little variation. As a result, $\Sigma_{\rm
  HI}$ transitions from being a relatively unimportant driver for SF in
the inner parts of galaxies to a key quantity at large radii. This is
apparent from the right panel of Figure \ref{fig4}, where we use
extremely deep FUV data from GALEX and \hi\ data from THINGS to trace SF and
\hi\ out to almost four optical radii in the nearby spiral M83
(\cite{bigiel10b}). The \hi\ depletion time (\hi-to-SFR ratio)
inferred from the radial profiles in this plot is approximately
constant at large radii: it is about a Hubble time, i.e., much longer
than the $\sim2$\,Gyr molecular gas depletion time scale observed in
the inner parts of galaxies. This suggests relatively fixed conditions
for molecular cloud formation and that assembling \htwo, rather than
forming stars out of \htwo, is the rate-limiting process for SF in
outer disks.

\section{The Composite Star Formation Law}

The combined distribution (inner and outer disks) in the left panel of
Figure \ref{fig4} can be divided into different parts according to gas
column density, each part describing the relation between gas and SFR
in a particular regime in a typical spiral galaxy disk. For low gas
columns (outer disks), the SFR scales with gas column, though with
significant scatter. At smaller radii and higher gas columns
--- corresponding to much of the area inside the star-forming disk ---
the distribution becomes much steeper. In this regime, knowing
$\Sigma_{\rm gas}$ alone is not enough to predict $\Sigma_{\rm SFR}$
with any accuracy. Across this regime the \htwo-to-\hi\ ratio is varying
steadily as a function of other environmental quantities. At yet smaller radii
and higher gas columns, the dominant phase of the ISM transitions
from atomic to molecular and a strong correlation emerges between
$\Sigma_{\rm gas}$ and $\Sigma_{\rm SFR}$.

There is good observational evidence (e.g., \cite{kennicutt98,gao04,greve05,bouche07})
that at higher gas columns, the relation steepens further, so that the
SFR-per-\htwo\ ratio is higher in starburst galaxies than in normal galaxy
disks. This may drive the frequent observation of $N\approx1.5$ in
starburst galaxies. However, it must be emphasized that there is
currently a lack of data probing normal disk galaxies and starbursts
in a self-consistent way, so the details at the high end of this
relation remain uncertain.

\section{Conclusions}

With vast improvements in the data available for nearby galaxies some
consensus is beginning to emerge on how different parts of galaxies populate the
$\Sigma_{\rm SFR}$-$\Sigma_{\rm gas}$ parameter space. The role of
environmental quantities other than gas surface density alone are
beginning to become clear and different relations are emerging for
different types and parts of galaxies. When viewed in detail the composite
relation may not be a simple power law, but it contains key
information to constrain theories and to benchmark simulations.

Challenges remain, too. The determination of star formation rates at
low surface brightness is still difficult. The use of CO to trace
\htwo\ underpins almost all of this work but the CO-to-\htwo\ conversion
factor remains imprecisely calibrated as a function of
environment. Finally, the fundamental units of star-formation,
individual molecular clouds, remain largely observationally inaccessible beyond
the Local Group --- a situation that will not change until ALMA begins
its full operations.

\acknowledgments
We thank the SOC and LOC for organizing this stimulating and productive meeting.
F.B. gratefully acknowledges financial support from the IAU. We thank
Andreas Schruba for providing us with the SFR-\htwo\ plot before publication.

\end{document}